\documentclass[11pt]{article}

\usepackage{hyperref}
\usepackage[utf8]{inputenc} 
\usepackage[T1]{fontenc}    
\usepackage{url}            
\usepackage{booktabs}       
\usepackage{amsfonts}       
\usepackage{nicefrac}       
\usepackage{microtype}      
\usepackage{amsmath,amssymb,subfigure,epsfig,bbm,mathrsfs,mathrsfs,dsfont, upgreek}
\usepackage{subfigure}
\usepackage{amstext}
\usepackage{array,xcolor}
\usepackage{color,soul}
\usepackage{hyperref}
\usepackage{caption}
\usepackage[all,cmtip]{xy}
\usepackage{makecell}
\usepackage{diagbox}
\usepackage{rotating}
\usepackage{multirow,bigdelim}
\usepackage{makeidx,overpic}
\usepackage{mdframed,algorithmic}
\usepackage[lined,boxruled]{algorithm2e}
\usepackage{boldline}

\newcommand{\R}{\mathbb{R}}
\newcommand{\C}{\mathbb{C}}

\newcommand{\E}{\mathbb{E}}

\newcommand{\vct}[1]{\boldsymbol{#1}}
\newcommand{\mtx}[1]{\boldsymbol{#1}}

\newcommand{\<}{\langle}
\renewcommand{\>}{\rangle}
\newcommand{\T}{\mathrm{T}}

\newcommand{\trace}{\operatorname{Tr}}

\newcommand{\set}[1]{\mathcal{#1}}

\DeclareMathOperator*{\minimize}{\text{minimize}}

\DeclareMathOperator*{\sto}{\mbox{subject to}}

\newcommand{\va}{\vct{a}}
\newcommand{\vb}{\vct{b}}
\newcommand{\vc}{\vct{c}}

\newcommand{\vh}{\vct{h}}

\newcommand{\vm}{\vct{m}}

\newcommand{\vu}{\vct{u}}
\newcommand{\vv}{\vct{v}}
\newcommand{\vw}{\vct{w}}
\newcommand{\vx}{\vct{x}}
\newcommand{\vy}{\vct{y}}
\newcommand{\vz}{\vct{z}}
\newcommand{\mA}{\mtx{A}}
\newcommand{\mB}{\mtx{B}}
\newcommand{\mC}{\mtx{C}}

\newcommand{\mF}{\mtx{F}}
\newcommand{\mG}{\mtx{G}}
\newcommand{\mH}{\mtx{H}}
\newcommand{\mI}{\mtx{I}}

\newcommand{\mM}{\mtx{M}}

\newcommand{\mP}{\mtx{P}}

\newcommand{\mU}{\mtx{U}}

\newcommand{\mW}{\mtx{W}}
\newcommand{\mX}{\mtx{X}}

\newcommand{\mZ}{\mtx{Z}}
\newcommand{\setA}{\set{A}}

\newcommand{\setH}{\set{H}}
\newcommand{\setI}{\set{I}}

\newcommand{\setL}{\set{L}}

\newcommand{\setN}{\set{N}}

\newcommand{\setP}{\set{P}}
\newcommand{\setQ}{\set{Q}}
\newcommand{\setR}{\set{R}}
\newcommand{\setS}{\set{S}}

\newcommand{\yl}{y_\ell}

\newcommand{\bl}{\vb_\ell}

\newcommand{\cl}{\vc_\ell}

\newcommand{\dM}{\delta \mM}
\newcommand{\dH}{\delta \mH}

\newcommand{\blt}{\bl^*}
\newcommand{\clt}{\cl^*}

\newcommand{\PP}{\mathbb{P}}

\newcommand{\Th}{T_{\tilde{h}}}
\newcommand{\Tm}{T_{\tilde{m}}}
\newcommand{\Thp}{T_{\tilde{h}}^\perp}
\newcommand{\Tmp}{T_{\tilde{m}}^\perp}
\newcommand{\PTh}{\setP_{\Th}}
\newcommand{\PTm}{\setP_{\Tm}}

\newtheorem{lemma}{Lemma}

\newtheorem{theorem}{Theorem}

\newcommand{\ba}[1]{{\va_{{#1},\ell}}}
\newcommand{\ind}{\mathbb{I}}
\newcommand{\balpha}{\boldsymbol{\alpha}}
\DeclareMathOperator*{\argmin}{arg\,min}

\begin{document}

\title{\vspace{-2cm}\bf{Blind Deconvolutional Phase Retrieval via Convex Programming}}
\author{Ali Ahmed\thanks{Department of Electrical Engineering, Information Technology University, Lahore. Email: {\tt ali.ahmed@itu.edu.pk}} \and  Alireza Aghasi\thanks{School of Business, Georgia State University, Atlanta, GA. Email: {\tt aaghasi@gsu.edu}} \and Paul Hand\thanks{Department of Computational and Applied Mathematics, Rice University, Houston, TX Email: {\tt hand@rice.edu}}}

\date{}    


\maketitle

\begin{abstract}
We consider the task of recovering two real or complex $m$-vectors from phaseless Fourier measurements of their circular convolution.  Our method is a novel convex relaxation that is based on a lifted matrix recovery formulation that allows a nontrivial convex relaxation of the bilinear measurements from convolution.    We prove that if  the two signals belong to known random subspaces of dimensions $k$ and $n$, then they can be recovered up to the inherent scaling ambiguity with $m  >> (k+n) \log^2 m$  phaseless measurements.  Our method provides the first theoretical recovery guarantee for this problem by a computationally efficient algorithm and does not require a solution estimate to be computed for initialization. Our proof is based Rademacher complexity estimates.  Additionally, we provide an ADMM implementation of the method and provide numerical experiments that verify the theory.
\end{abstract}

\section{Introduction}
This paper considers recovery of two unknown signals (real- or complex-valued) from the magnitude only measurements of their convolution.  Let $\vw$, and $\vx$ be vectors residing in $\setH^m$, where $\setH$ denotes either $\R$, or $\C$. Moreover, denote by $\mF$ the DFT matrix with entries $F[\omega,t] =\tfrac{1}{\sqrt{m}} \mathrm{e}^{-j2\pi \omega t/\sqrt{m}}, ~ 1 \leq \omega, t \leq m.$ We observe the phaseless Fourier coefficients of the circular convolution $\vw \circledast \vx$ of $\vw$, and $\vx$ 
\begin{align}\label{eq:measurements}
\vy = |\mF(\vw \circledast \vx)|,
\end{align}
where $|\vz|$ returns the element wise absolute value of the vector $\vz$. We are interested in recovering $\vw$, and $\vx$ from the phaseless measurements $\vy$ of their circular convolution. In other words, the problem concerns blind deconvolution of two signals from phaseless measurements. The problem can also be viewed as identifying the structural properties on $\vw$ such that its convolution with the signal/image of interest $\vx$ makes the phase retrieval of a signal $\vx$ well-posed. 
Since $\vw$, and $\vx$ are both unknown, and in addition, the measurements are phaseless, the inverse problem becomes severly ill-posed as many pairs of $\vw$, and $\vx$ correspond to the same $\vy$. We show that this non-linear problem can be efficiently solved, under Gaussian measurements, using a semidefinite program and also theoretically prove this assertion. We also propose a heuristic approach to solve the proposed semidefinite program computationally efficiently. Numerical experiments show that, using this algorithm, one can successfully recover a blurred image from the magnitude only measurements of its Fourier spectrum. 

Phase retrieval has been of continued interest in the fields of signal processing, imaging, physics, computational science, etc. Perhaps, the single most important context in which phase retrieval arises is the X-ray crystallography \cite{harrison1993phase,millane1990phase}, where the far-field pattern of X-rays scattered from a crystal form a Fourier transform of its image, and it is only possible to measure the intensities of the electromagnetic radiation. However, with the advancement of imaging technologies, the phase retrieval problem continues to arise in several other imaging modalities such as diffraction imaging \cite{bunk2007diffractive}, microscopy \cite{miao2008extending}, and astronomical imaging\cite{fienup1987phase}. In the imaging context, the result in this paper would mean that if rays are convolved with a \textit{generic} pattern (either man made or naturally arising due to propagation of light through some unknown media) $\vw$ prior to being scattered/reflected from the object, the image of the object can be recovered from the Fourier intensity measurements later on. As is well known from Fourier optics \cite{goodman2008introduction}, the convolution of a visible light with a generic pattern can be implemented using a lens-grating-lens setup. 

Blind deconvolution is a fundamental problem in signal processing, communications, and in general system theory. Visible light communication has been proposed as a standard in 5G communications for local area networks \cite{azhar2013gigabit,retamal20154,azhar2010demonstration}. Propagation of information carrying light through an unknown communication medium is modeled as a convolution. The channel is unknown and at the receiver it is generally difficult to measure the phase information in the propagated light. The result in this paper says that the transmitted signal can be blindly deconvolved from the unknown channel from the Fourier intensity measurements of the light only. The reader is referred to Section \ref{sec:vis} of the Appendix for a detailed description of the visible light communication and its connection to our formulation. 

\subsection{Observations in Matrix Form}

The phase retrieval, and blind deconvolution problem has been extensively studied in signal processing community in recent years \cite{candes2015phase,ahmed2014blind} by lifting the unknown vectors to a higher dimensional matrix space formed by their outer products. The resulting rank-1 matrix is recovered using nuclear norm as a convex relaxation of the non-convex rank constraint. Recently, other forms of convex relaxations have been proposed \cite{bahmaniphaseretrieval,goldstein2018phasemax,aghasi2017branchhull,aghasi2017convex} that solve both the problems in the native (unlifted) space leading to computationally efficiently solvable convex programs. This paper handles the non-linear convolutional phase retrieval problem by lifting it into a bilinear problem. The resulting problem, though still non-convex, gives way to an effective convex relaxation that provably recovers $\vw$, and $\vx$ exactly. 

It is clear from \eqref{eq:measurements} that uniquely recovering $\vw$, and $\vx$ is not possible without extra knowledge or information about the problem. We will address the problem under a broad and generally applicable structural assumptions that both the vectors $\vw$, and $\vx$ are members of known subspaces of $\setH^m$. This means that $\vw$, and $\vx$ can be parameterized in terms of unknown lower dimensional vectors $\vh \in \setH^k$, and $\vm \in \setH^n$, respectively as follows
\begin{align}\label{eq:subspace-constraints}
\vw = \mB\vh, \ \vx = \mC \vm, 
\end{align}
where $\mB \in \setH^{m \times k}$, and $\mC \in \setH^{m \times n}$ are known matrices whose columns span the subspaces in which $\vw$, and $\vx$ reside, respectively. Recovering $\vh$, and $\vm$ would imply the recovery of $\vw$, and $\vx$, therefore, we take $\vh$, and $\vm$ as the unknowns in the inverse problem henceforth. 
Since the circular convolution operator diagonalizes in the Fourier domain, the measurements in \eqref{eq:measurements} take the following form after incorporating the subspace constraints in \eqref{eq:subspace-constraints} 
\begin{align*}
\vy = \tfrac{1}{\sqrt{m}}|\hat{\mB}\vh \odot \hat{\mC}\vm|,
\end{align*}
where $\hat{\mB} = \sqrt{m}\mF\mB$, $\hat{\mC} = \sqrt{m}\mF\mC$, and $\odot$ represent the Hadamard product. Denoting by $\bl$ and $\cl$ the rows of $\hat{\mB}$, and $\hat{\mC}$, respectively, the entries of the measurements $\vy$ can be expressed as
\begin{align*}
y^2_\ell =\tfrac{1}{m} |\<\bl,\vh\>\<\cl,\vm\>|^2,\ \ell = 1,2,3,\ldots, m.
\end{align*}
Evidently the problem is non-linear in both unknowns. However, it reduces to a bilinear problem in the lifted variables $\vh\vh^*$, and $\vm\vm^*$ taking the form 
\begin{align}\label{eq:lifted-measurements}
y^2_\ell = \tfrac{1}{m}\<\bl\blt,\vh\vh^*\>\<\cl\clt,\vm\vm^*\> = \tfrac{1}{m}\<\bl\blt,\mH\>\<\cl\clt,\mM\>,
\end{align}
where $\mH$, and $\mM$ are the rank-1 matrices $\vh\vh^*$, and $\vm\vm^*$, respectively. Treating the lifted variables $\mH$, and $\mM$ as unknowns makes the measurements bilinear in the unknowns; a structure that will help us formulate an effective convex relaxation. 
\subsection{Novel Convex Relaxation}

The task of recovering $\mH$, and $\mM$ from $\vy$ in \eqref{eq:lifted-measurements} can be naturally posed as an optimization program 
\begin{align}\label{eq:raw-optimization-program}
&\text{find} ~ \mH, \mM \\
&\text{subject to} ~ \tfrac{1}{m}\< \bl\blt,\mH\>\< \cl\clt,\mM\> = y^2_\ell , ~ \ell = 1,2,3, \ldots, m.\notag\\
& \qquad\qquad \qquad \qquad \text{rank}(\mH) = 1, ~ \text{rank}(\mM) = 1.\notag
\end{align}
However, both the measurement and the rank constraints are non-convex.  Further, the immediate convex relaxation of each measurement constraint is trivial, as the convex hull of the set of $(\mH, \mM)$ satisfying $y^2_\ell =  \tfrac{1}{m}\< \bl\blt,\mH\>\< \cl\clt,\mM\>$ is the set of all possible $(\mH, \mM)$.

\begin{figure}
\begin{overpic}[ width=0.57\textwidth,height=0.3\textwidth,tics=10]{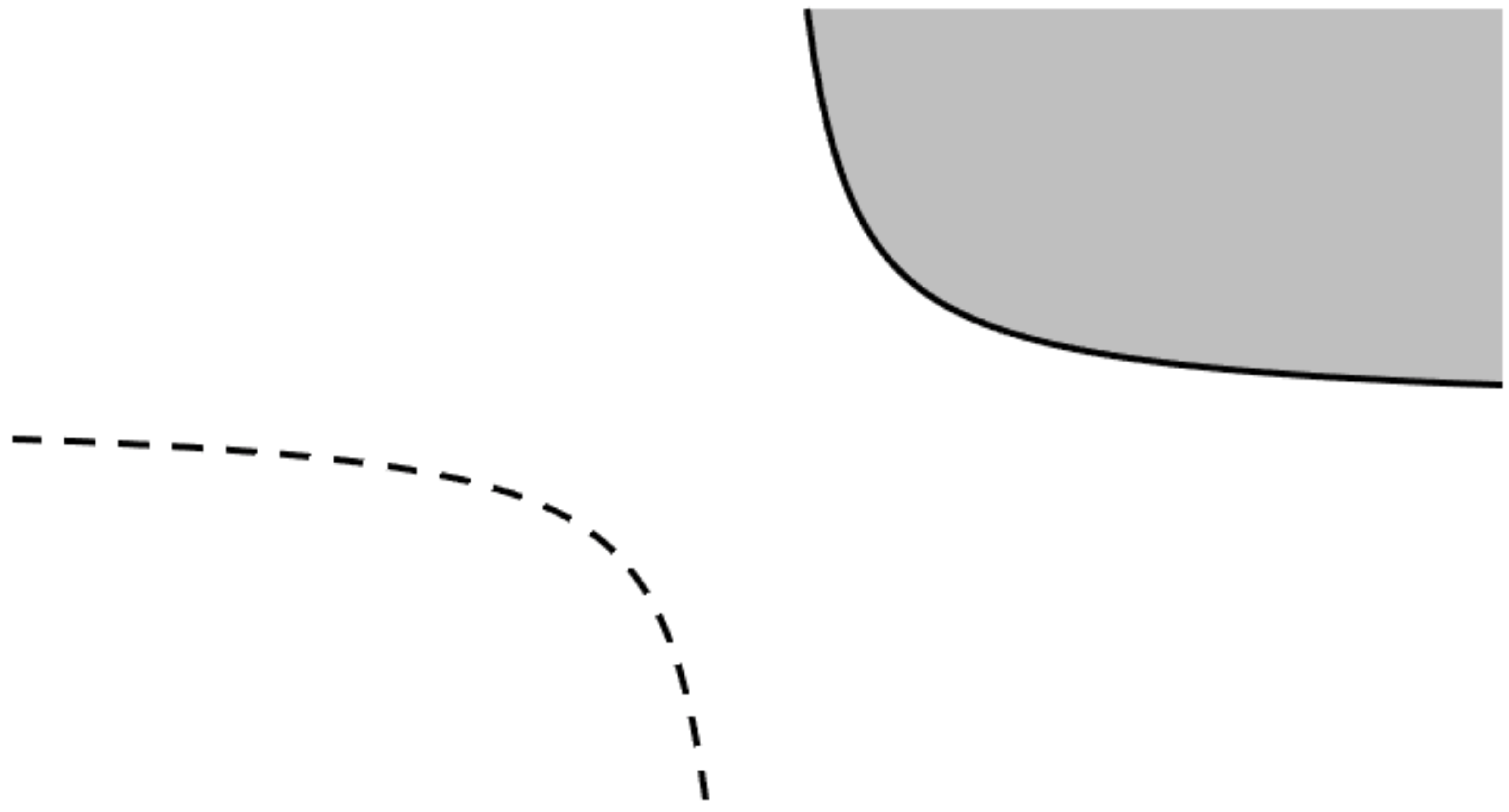}
\linethickness{0.3pt}
\put(-1,25.8){\color{black}\line(1,0){103}}
\put(50,.5){\color{black}\line(0,1){52.5}}
\put(47,21){$0$}
\put(43.5,52){$v_\ell$}
\put(100,21){$u_\ell$}
\put(51,34){\rotatebox{-30}{\scalebox{.75}{$\tfrac{1}{m}u_\ell v_\ell =y_\ell^2 $}}}
\put(56,41.5){\rotatebox{0}{\scalebox{.65}{$\mbox{Conv}\!\left\{\begin{pmatrix}u_\ell\\v_\ell\end{pmatrix}: \tfrac{1}{m} u_\ell v_\ell =y_\ell^2, u_\ell>0\right\}$}}}
 \end{overpic}
\hspace{.7cm}\includegraphics[trim = 10cm 5cm 5cm 0cm,clip,scale = 0.35]{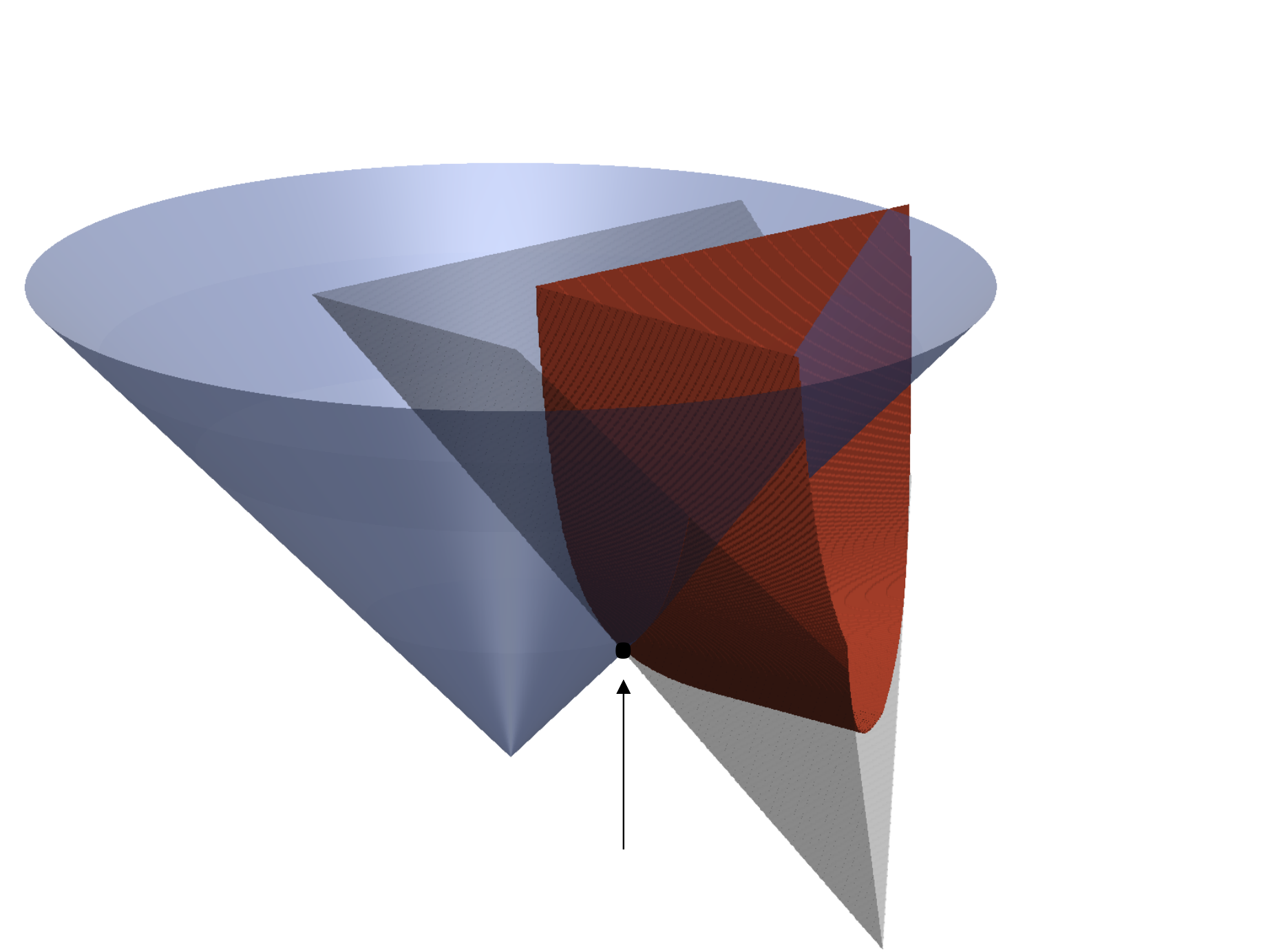}
	\caption{Left: Restriction of the hyperbolic constraint to the first quadrant;  Right: Abstract Illustration of the Geometry of the Convex Relaxation. PSD cone (blue) and the surface of the hyperbolic set (red) formed by two intersecting hyperbolas $(m=2)$. Evidently, there are multiple points on the surface and also in the convex hull of the hyperbolic set that lie on the PSD cone. The minimizer of the optimization program \eqref{eq:convex-optimization-program} picks the one with minimum trace that happens to lie at the intersection of hyperbolic ridge and the PSD cone (pointed out by an arrow). The gray envelope of two $(m =2)$ hyperplanes  surrounding the hyperbolic set correspond to the linearization of the hyperbolic set at the minimizer; this forms the basis of a connected linearly constrained program later in \eqref{eq:linearly-constrained-program}.}
	\label{fig:Geometry}
\end{figure}
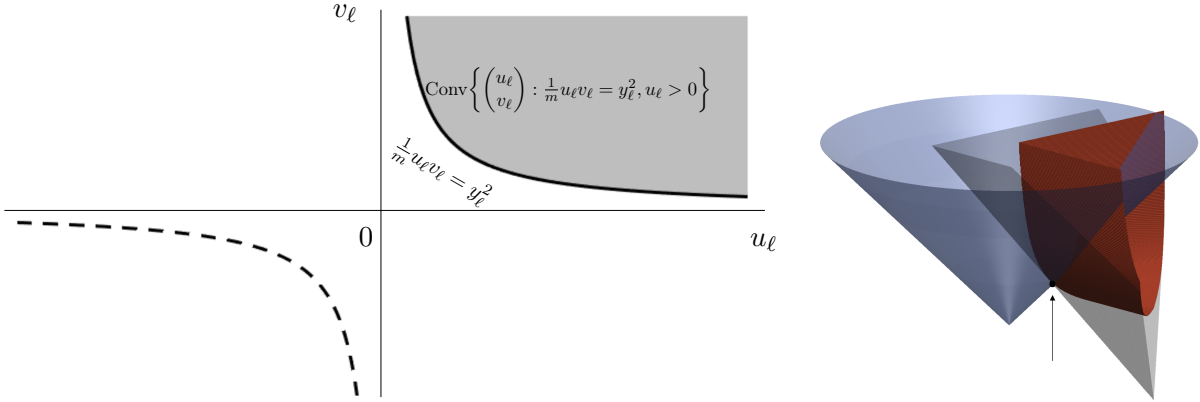

To derive our convex relaxation, 
    recall that the true $\mH = \vh\vh^*$, and $\mM = \vm\vm^*$ are also positive semidefinite (PSD). This means that incorporating the PSD constraint in the optimization program translates into the fact that the variables $u_\ell = \< \bl\blt,\mH\>$ and $v_\ell = \< \cl\clt,\mM\>$ are necessarily non-negative. That is, 
\begin{align*}
\mH \succcurlyeq \mathbf{0}, \ \text{and} \  \mM \succcurlyeq \mathbf{0}  \implies u_\ell \geq 0, \ \text{and} \ v_\ell \geq 0,
\end{align*}
where the implication simply follows by the definition of PSD matrices. This observation restricts the hyperbolic constraint set in Figure~\ref{fig:Geometry} to the first quadrant only. For a fixed $\ell$, we propose replacing the non-convex hyperbolic set $\{(u_\ell, v_\ell) \in \R^2\ |\ \tfrac{1}{m} u_\ell v_\ell = y^2_\ell, u_\ell \geq 0 , \ v_\ell \geq 0 \}$ with its convex hull $\{(u_\ell, v_\ell) \in \R^2 \ |\ \tfrac{1}{m} u_\ell v_\ell \geq y^2_\ell, u_\ell \geq 0 , \ v_\ell \geq 0 \}.$  In short, our convex relaxation is possible because the PSD constraint from lifting happens to select a specific branch of the hyperbola given by any particular bilinear measurement, and this single branch has a nontrivial convex hull.

The rest of the convex relaxation is standard, as the rank constraint in \eqref{eq:raw-optimization-program} is then relaxed with a nuclear-norm minimization, which reduces to trace minimization in the PSD case. Hence, we study the convex  program 
\begin{align}\label{eq:convex-optimization-program}
&\minimize   \ \trace(\mH)+\trace(\mM) \\
&\text{subject to} \ \tfrac{1}{m}\<\bl\blt, \mH\>\<\cl\clt,\mM\> \geq y^2_\ell, \ \ell = 1,2, \ldots, m\notag\\
&\qquad\qquad  \mH \succcurlyeq \mathbf{0}, \  \mM \succcurlyeq \mathbf{0}\notag.
\end{align}

\subsection{Main Result}
As we are presenting the first analytical results on this problem, we choose the subspace matrices $\mB$, and $\mC$ to be standard Gaussian:
 \begin{align}\label{eq:BC-Random-Model}
 & B[\ell,i] \sim \text{Normal}(0,\tfrac{1}{m}), (\ell,i) \in [m] \times [k], \text{and} \ C[\ell,i] \sim \text{Normal}(0,\tfrac{1}{m}), (\ell,i) \in [m] \times [n].
 \end{align}
 Note that this choice results in $\bl,\cl \sim \text{Normal}(\mathbf{0},\mI)$. We show that with this choice the optimization program in \eqref{eq:convex-optimization-program} recovers a global scaling of $(\alpha \mH^\natural, \alpha^{-1}\mM^\natural)$ of the true solution $(\mH^\natural, \mM^\natural).$ We will interchangeably use the notation $(\mH,\mM)\in (\setH^{k\times k}, \setH^{n \times n})$ to denote the pair of matrices $\mH$ and $\mM$, or the block diagonal matrix
 \begin{align}\label{eq:matrix-pair}
 (\mH,\mM) = \begin{bmatrix} \mH & \mathbf{0}\\
 \mathbf{0} & \mM
 \end{bmatrix}.
 \end{align}
The exact value of the unknown scalar multiple $\alpha$ can be characterized for the solution of \eqref{eq:convex-optimization-program}. Observe that the solution $(\widehat{\mH},\widehat{\mM})$ of the convex optimization program in \eqref{eq:convex-optimization-program} obeys $\trace(\widehat{\mH}) = \trace(\widehat{\mM})$. We aim to show that the solution of the optimization program recovers the scaling $(\tilde{\mH}, \tilde{\mM})$ of the true solution $(\mH^\natural,\mM^\natural)$:
 \begin{align*}
 \tilde{\mH} = \sqrt{\frac{\trace(\mM^\natural)}{\trace(\mH^\natural)}} \mH^\natural, ~ \tilde{\mM} = \sqrt{\frac{\trace(\mH^\natural)}{\trace(\mM^\natural)}} \mM^\natural.
 \end{align*}
 Note that $\trace(\tilde{\mH}) = \trace(\tilde{\mM})$. 
 The main result can now be stated as follows. 
 \begin{theorem}[Exact Recovery]\label{thm:main-theorem}
 	Given the magnitude only spectrum measurements \eqref{eq:measurements} of the convolution of two unknown vectors $\vw^\natural$, and $\vx^\natural$ in $\setH^m$. Suppose that $\vw^\natural$, and $\vx^\natural$ are generated as in \eqref{eq:subspace-constraints}, where $\mB$, and $\mC$ are known standard Gaussian matrices as in \eqref{eq:BC-Random-Model}. Then the convex optimization program in \eqref{eq:convex-optimization-program} uniquely recovers $(\alpha \mH^\natural, \alpha^{-1}\mM^\natural)$ for $\alpha = \sqrt{\frac{\trace \mM^\natural}{\trace \mH^\natural}}$ 
	with probability at least $1-\exp(-\tfrac{1}{2} mt^2)$ whenever $m \geq c_t (k+n)\log^2 m$,
 	where $c_t$ is a constant that depends on $t \geq 0$. 
 \end{theorem}
\subsection{Main Contributions}
In this paper, we study the combination of two important and notoriously challenging signal recovery problems: phase retrieval and blind deconvolution.  We introduce a novel convex formulation that is possible because the algebraic structure from lifting resolves the bilinear ambiguity just enough to permit a nontrivial convex relaxation of the measurements.  The strengths of our approach are that it allows a novel convex program that is the first to provably permit recovery guarantees with optimal sample complexity for the joint task of phase retrieval and blind deconvolution when the signals belong to known random subspaces.  Additionally, unlike many recent convex relaxations and nonconvex approaches, our approach does not require an initialization or estimate of the true solution in order to be stated or solved.  Admittedly, our method, directly interpreted, is computationally prohibitive for large problem sizes because lifting squares the dimensionality of the problem.  Nonetheless, techniques, such as Burer-Monteiro approaches that only maintain low-rank representations \cite{Burer2003}, have been developed for similar problems.  This current work provides the theoretical justification for the exploration of such problems in this difficult combination of phase retrieval and blind deconvolution, and we leave such work for future research.

We do not want to give the reader the impression that the present paper solves the problem of blind deconvolutional phase retrieval in practice.  The numerical experiments we perform do indeed show excellent agreement with the theorem in the case of random subspaces.  Such subspaces are unlikely to appear in practice, and typically appropriate subspaces would be deterministic, including partial Discrete Cosine Transforms or partial Discrete Wavelet Transforms.  Numerical experiments, not shown, indicate that our convex relaxation is less effective for the cases of these deterministic subspaces.  We suspect this is due to the fact that the subspaces for both measurements should be mutually incoherent, in addition to both being incoherent with respect to the Fourier basis given by the measurements.  As with the initial recovery theory for the problems of compressed sensing and phase retrieval, we have studied the random case in order to show information theoretically optimal sample complexity is possible by efficient algorithms.  Based on this work, it is clear that blind deconvolutional phase retrieval is still a very challenging problem in the presence of deterministic matrices, and one for which development of convex or nonconvex methods may provide substantial progress in applications.

\section{Proof of Theorem \ref{thm:main-theorem}}

To prove Theorem \ref{thm:main-theorem}, we will show that $(\tilde{\mH},\tilde{\mM})$ is the unique minimizer of an optimization program with a larger feasible set defined by linear constraints.
\begin{lemma}\label{lem:Hyp-to-Lin}
	If $(\tilde{\mH},\tilde{\mM})$ is the unique solution to 
	\begin{align}\label{eq:linearly-constrained-program}
	&\minimize ~ \|\mH\|_* + \|\mM\|_* \\
	&\sto  \ \tfrac{1}{m} (\< \bl\blt,\mH\>\< \cl\clt,\tilde{\mM}\>+\< \bl\blt,\tilde{\mH}\>\< \cl\clt,\mM\>) \geq  2y^2_\ell, \notag \\
	& \qquad\qquad \qquad ~ \ell = 1,2,3, \ldots, m. \notag 
	\end{align}
	then $(\tilde{\mH},\tilde{\mM})$ is the unique solution to \eqref{eq:convex-optimization-program}. 
\end{lemma}
\textbf{proof:}\\
	Start by observing that the trace in \eqref{eq:convex-optimization-program} can be replaced with nuclear norm as on the set of PSD matrices both are equivalent. This gives 
	\begin{align}\label{eq:nuc-norm-min}
	&\minimize   \ \|\mH\|_*+\|\mM\|_* \\
	&\text{subject to} \ \tfrac{1}{m}\<\bl\blt, \mH\>\<\cl\clt,\mM\> \geq y^2_\ell, \ \ell = 1,2, \ldots, m\notag\\
	&\qquad\qquad  \mH \succcurlyeq \mathbf{0}, \  \mM \succcurlyeq \mathbf{0}\notag.
	\end{align}

	It suffices now to show that the feasible set of \eqref{eq:linearly-constrained-program} contains the feasible set of \eqref{eq:nuc-norm-min}. Recall the notations 
	\[
	u_\ell = \< \bl\blt,\mH\>,  ~v_{\ell} = \< \cl\clt,\mM\>,  ~\tilde{u}_\ell = \< \bl\blt,\tilde{\mH}\>, ~\text{and} ~\tilde{v}_{\ell} = \< \cl\clt,\tilde{\mM}\>.
	\]
	Using the fact that a convex set with smooth boundary is contained in a half space defined by the tangent hyperplane at any point on the boundary of the set. Consider the point $(\tilde{u}_\ell,\tilde{v}_\ell) \in \R^2$, and observe that 
	\begin{align*}
	&\left\{ (u_\ell,v_\ell) \in \R^2 ~ | ~ \tfrac{1}{m}u_\ell v_\ell \geq y^2_\ell, u_\ell \geq 0, \ \text{and} \ v_\ell \geq 0 \right\}\subseteq \left\{ (u_\ell,v_\ell) \in \R^2 ~ | ~ \tfrac{1}{m}\begin{bmatrix} \tilde{v}_\ell \\ \tilde{u}_\ell \end{bmatrix} \cdot \begin{bmatrix} u_\ell - \tilde{u}_\ell \\ v_\ell - \tilde{v}_\ell \end{bmatrix} \geq 0 \right\}.
	\end{align*}
Rewriting $u_\ell$ and $v_\ell$ in the form of original constraints, we have that any feasible point $(\tilde{\mH},\tilde{\mM})$ of \eqref{eq:nuc-norm-min} satisfies $\tfrac{1}{m}(\< \bl\blt,\mH\>\< \cl\clt,\tilde{\mM}\>+\< \bl\blt,\tilde{\mH}\>\< \cl\clt,\mM\>) \geq  2y^2_\ell, ~\ell = 1,2,3, \ldots, m.$ $\square$

The geometry of the linearly constrained program \eqref{eq:linearly-constrained-program} is also shown in Figure \ref{fig:Geometry} (Right), where the hyperbolic set is replaced by an envelop of hyperplanes defined by the linear constraints of \eqref{eq:linearly-constrained-program}. Visually it is clear from Figure \ref{fig:Geometry} that the feasible set of \eqref{eq:linearly-constrained-program} is larger than that of \eqref{eq:convex-optimization-program}. 

Define a set $\setS: = \{ (\mH,\mM) ~| ~ (\mH,\mM) = \alpha (-\tilde{\mH},\tilde{\mM}), ~ \text{and} ~ \alpha \in [-1,1]\}$, and $\mA_{\ell} = (\tilde{v}_\ell\bl\blt ,\tilde{u}_\ell \cl\clt ) \in \setH^{(k+n) \times (k+n)},$ and define a linear map $\setA: \setH^{(k+n) \times (k+n)} \rightarrow \setH^m$ as $$
\setA((\mH,\mM)) = [\<\mA_1,(\mH,\mM)\>, \ldots, \<\mA_m,(\mH,\mM)\>]^\T;$$ 
one can imagine $\setA$ as a matrix with vectorized $\mA_\ell$ as its rows. The linear constraints in the \eqref{eq:linearly-constrained-program} are $\setA((\mH,\mM)) \geq 2 \vy^2$; the inequality here applies elementwise. Furthermore, define $\setN : = \text{span}((-\tilde{\mH},\tilde{\mM}))$, and it is easy to see that $\setS \subset \setN \subseteq \text{Null}(\setA).$

We want to show that any feasible perturbation $(\dH,\dM)$ around the truth $(\tilde{\mH},\tilde{\mM})$ strictly increases the objective. From the discussion above, it is clear that the perturbations $(\dH,\dM) \in \setS$ do not change the objective and also lead to feasible points of \eqref{eq:linearly-constrained-program}. Our general strategy will be to resolve any perturbation $(\dH,\dM)$ into two components, one in $\setN$ and the other in $\setN_\perp$, where $\setN_{\perp}$ is the orthogonal complement of the subspace $\setN$.  The component in $\setN$ does not affect the objective. We show that the components in $\setN_\perp$ of all the feasible perturbations lead to a strict increase in the objective of \eqref{eq:linearly-constrained-program}. This should imply that  that the minimizer of \eqref{eq:linearly-constrained-program}  can be anywhere in the set $(\tilde{\mH},\tilde{\mM}) \oplus \setN$. However, as we are minimizing the (trace) norms, an arbitrary large scaling of the solution is prevented and it is restricted to the subset $(\tilde{\mH},\tilde{\mM}) \oplus \setS$. Moreover, among these solutions only $(\tilde{\mH},\tilde{\mM})$ lies in the feasible set of \eqref{eq:nuc-norm-min}. Given this and the fact that $(\tilde{\mH}, \tilde{\mM})$ is a minimizer of \eqref{eq:linearly-constrained-program} implies that $(\tilde{\mH}, \tilde{\mM})$ is the unique minimizer of \eqref{eq:nuc-norm-min}. 

We begin by characterizing the set of descent directions for the objective function of the optimization program \eqref{eq:linearly-constrained-program}. Let $\Th$, and $\Tm$ be the set of symmetric matrices of the form 
\begin{align*}
\Th := \{ \mX = \tilde{\vh}\vz^* + \vz\tilde{\vh}^*\}, ~ \Tm := \{ \mX = \tilde{\vm}\vz^* + \vz\tilde{\vm}^*\},
\end{align*}
and denote the orthogonal complements by $\Thp$, and $\Tmp$, respectively. Note that $\mX \in \Th^\perp$ iff both the row and column spaces of $\mX$ are perpendicular to $\tilde{\vh}$. $\PTh$ denotes the orthogonal projection onto the set $\Th$, and a matrix $\mX$ of appropriate dimensions can be projected into $\Th$ as
\[
\PTh(\mX) : = \tfrac{\tilde{\vh}\tilde{\vh}^* }{\|\tilde{\vh}\|_2^2} \mX+ \mX \tfrac{\tilde{\vh}\tilde{\vh}^* }{\|\tilde{\vh}\|_2^2}  - \tfrac{\tilde{\vh}\tilde{\vh}^* }{\|\tilde{\vh}\|_2^2}\mX\tfrac{\tilde{\vh}\tilde{\vh}^* }{\|\tilde{\vh}\|_2^2}
\]
 Similarly, define the projection operator $\PTm$. The projection onto orthogonal complements are then simply $\setP_{\Thp} : = \setI - \PTh$, and  $\setP_{\Tmp}: = \setI - \PTm$, where $\setI$ is the identity operator. We use $\mX_{\Th}$ as a shorthand for $\PTh(\mX)$. Using the notation in \eqref{eq:matrix-pair}, the objective of \eqref{eq:linearly-constrained-program} is $\|(\mH,\mM)\|_*$, and subgradient of the objective at the proposed solution $(\tilde{\mH}, \tilde{\mM})$ is 
\begin{align*}
\partial \|(\tilde{\mH},\tilde{\mM})\|_* :=\big\{\mG = (\tilde{\vh}\tilde{\vh}^*,\tilde{\vm}\tilde{\vm}^*) + (\mW_{\Thp}, \mW_{\Tmp}), ~ \|(\mW_{\Thp}, \mW_{\Tmp})\| \leq 1\big\}.
\end{align*}
The set $\setQ$ of descent directions of the objective of \eqref{eq:linearly-constrained-program} is defined as 
\begin{align}\label{eq:setQ}
&\big\{ (\dH,\dM) \in \setN_{\perp}: \big<(\mG, (\dH,\dM)\big\> \leq 0,  \forall \mG \in \partial \|(\tilde{\mH},\tilde{\mM})\|_*\big\}\subseteq \notag \\
&\big\{ (\dH,\dM)\in \setN_{\perp}: \big<(\tilde{\vh}\tilde{\vh}^*,\tilde{\vm}\tilde{\vm}^*),(\dH,\dM)\big> +\notag\\
&\qquad \qquad\qquad \qquad \| (\dH_{\Thp}, \dM_{\Tmp} )\|_* \leq 0, \forall \mG \in \partial \|(\tilde{\mH},\tilde{\mM})\|_*\big\}\subset  \notag\\
&\big\{ (\dH,\dM)\in \setN_{\perp}:  \| (\dH_{\Thp}, \dM_{\Tmp} )\|_* \leq  \| (\dH_{\Th}, \dM_{\Tm} )\|_F, ~  \forall \mG \in \partial \|(\tilde{\mH},\tilde{\mM})\|_* \big\} \notag \\
&=: \setQ.
\end{align}
We quantify the "width" of the set of descent directions $\setQ$ through a Rademacher complexity, and a probability that the gradients of the constraint functions of \eqref{eq:linearly-constrained-program} lie in a certain half space. This enables us to build an argument using the small ball method \cite{koltchinskii2015bounding,mendelson2014learning} that it is unlikely to have points that meet the constraints in \eqref{eq:linearly-constrained-program} and still be in $\setQ$. Before moving forward, we introduce the above mentioned Rademacher complexity and probability term.   

Denote the constraint functions as\footnote{For brevity, we will often drop the dependence on $\mH$, and $\mM$ in the notation $f_\ell(\mH,\mM)$} $
f_\ell(\mH,\mM) = \tilde{u}_\ell \< \cl\clt,\mM\> + \tilde{v}_\ell \<\bl\blt,\mH\>.$
For a set $\setQ \subset (\setH^{k \times k}, \setH^{n \times n})$, the Rademacher complexity of the gradients $\nabla f_{\ell}= (\tfrac{\partial f_{\ell} }{\partial \mH}, \tfrac{\partial f_{\ell} }{\partial \mM}) = (\tilde{v}_\ell\bl\blt, \tilde{u}_\ell\cl\clt )$ is defined as 
\begin{align}\label{eq:Rademacher-Complexity}
\mathfrak{C}(\setQ) := \E \sup_{(\mH,\mM) \in \setQ}\tfrac{1}{\sqrt{m}}\sum_{\ell=1}^m \varepsilon_{\ell}\left\< \nabla f_{\ell}, \tfrac{(\mH,\mM)}{\|(\mH,\mM)\|_F}\right\>,
\end{align}
where $\varepsilon_\ell, ~\ell = 1,2,3,\ldots, m$ are iid Rademacher random variables independent of everything else in the expression. For a convex set $\setQ$, $\mathfrak{C}(\setQ)$ is a measure of the width of $\setQ$ around origin interms of the gradients $\nabla f_\ell, ~ \ell = 1,2,3,\ldots, m$. For example, random choice of gradient might yield little overlap with a structured set $\setQ$ leading to a smaller complexity $\mathfrak{\setQ}$. 

Our result also depends on a probability $p_{\tau}(\setQ)$ and a positive parameter $\tau$ defined as 
\begin{align}\label{eq:p_tau}
p_{\tau}(\setQ) := \inf_{(\mH,\mM) \in \setQ} \mathbb{P}\big( \<\nabla f, (\mH, \mM)\> \geq \tau \|(\mH,\mM)\|_F \big).
\end{align}
The probability $p_{\tau}(\setQ)$ quantifies visibility of the set $\setQ$ through the gradient vectors $\nabla f$. A small value of $\tau$ and $p_{\tau}(\setQ)$ means that the set $\setQ$ mainly remains invisible through the lenses of $\nabla f_\ell, \ell = 1,2,3, \ldots, m$. This can be appreciated just by noting that $p_\tau(\setQ)$ depends on the correlation of the elements of $\setQ$ with the gradient vectors $\nabla f_\ell$. 

Following lemma shows that the minimizer of the linear program \eqref{eq:linearly-constrained-program} almost always resides in the desired set $(\tilde{\mH},\tilde{\mM})\oplus \setS$ for a sufficiently large $m$ quantified interms of $\mathfrak{C}(\setQ)$, $p_{\tau}(\setQ)$, and $\tau$. 
\begin{lemma}\label{lem:Sample-Complexity}
	Consider the optimization program in \eqref{eq:linearly-constrained-program} and $\setQ$, characterized in \eqref{eq:setQ}, be the set of descent directions for which $\mathfrak{C}(\setQ)$, and $p_{\tau}(\setQ)$ can be determined using \eqref{eq:Rademacher-Complexity} and \eqref{eq:p_tau}. Choose 
	\begin{align*}
	m \geq \left(\frac{2\mathfrak{C}(\setQ)+t\tau}{\tau p_{\tau(\setQ)}}\right)^2
	\end{align*}
	for any $t >0$. Then the minimizer $(\widehat{\mH}, \widehat{\mM})$ of \eqref{eq:linearly-constrained-program} lies in the set $(\tilde{\mH},\tilde{\mM}) \oplus \setS$
	with probability at least $1-\mathrm{e}^{-2mt^2}$. 
\end{lemma}
Proof of this lemma is based on small ball method developed in \cite{koltchinskii2015bounding,mendelson2014learning} and further studied in  \cite{lecue2018regularization,lecue2017regularization}. The proof is mainly repeated using the argument in \cite{bahmani2017anchored}, and is provided in the Appendix for completeness.

With Lemma \ref{lem:Sample-Complexity} in place, an application Lemma \ref{lem:Hyp-to-Lin} and the discussion after it proves that for choice of $m$ outlined in Lemma \ref{lem:Sample-Complexity}, $(\tilde{\mH},\tilde{\mM})$ is the unique minimizer of \eqref{eq:convex-optimization-program}. The last missing piece in the proof of Theorem \ref{thm:main-theorem} is the computation of the Rademacher complexity $\mathfrak{C}(\setQ)$, and $p_\tau(\setQ)$ for the $\setQ$. 

\subsection{Rademacher Complexity}
We begin with evaluation of the complexity $\mathfrak{C}(\setQ)$
\begin{align*}
\mathfrak{C}(\setQ) &:= \E \sup_{(\dH,\dM) \in \setQ} ~ \frac{1}{\sqrt{m}}  \sum_{\ell=1}^m \varepsilon_\ell \Big\< \nabla f_\ell, \tfrac{(\dH,\dM)}{\|(\dH,\dM)\|_F}\Big\>
\end{align*}
Splitting $(\dH,\dM)$ between $(\Th,\Tm)$, and $(\Th^\perp,\Tm^\perp)$, and using Holder's inequalities, we obtain
\begin{align*}
\mathfrak{C}(\setQ) & \leq \E \Big\| \frac{1}{\sqrt{m}}\sum_{\ell=1}^m\varepsilon_\ell ( \tilde{v}_\ell \PTh(\bl\blt),\tilde{u}_\ell \PTm(\cl\clt))
\Big\|_F \cdot \sup_{(\dH,\dM) \in \setQ}~ \left\| \tfrac{(\dH_{\Th},\dM_{\Tm})}{\|(\dH,\dM)\|_F}\right\|_F\\
&+ \E \Big\| \frac{1}{\sqrt{m}}\sum_{\ell=1}^m\varepsilon_\ell ( \tilde{v}_\ell\bl\blt,\tilde{u}_\ell \cl\clt)
\Big\|\cdot \sup_{(\dH,\dM) \in \setQ}~ \left\|\tfrac{ (\dH_{\Thp}, \dM_{\Tmp })}{\| (\dH, \dM )\|_F}\right\|_*
\end{align*}
On the set $\setQ$, defined in \eqref{eq:setQ}, we have 
\begin{align*}
\Big\|\tfrac{ (\dH_{\Thp}, \dM_{\Tmp })}{\| (\dH, \dM )\|_F}\Big\|_* \leq \Big\| \tfrac{(\dH_{\Th},\dM_{\Tm})}{\|(\dH,\dM)\|_F}\Big\|_F \leq 1.
\end{align*}
Using Jensen's inequality, the first expectation simply becomes
\begin{align*}
&\E\Big\| \frac{1}{\sqrt{m}}\sum_{\ell=1}^m\varepsilon_\ell \big( \tilde{v}_\ell \PTh(\bl\blt),\tilde{u}_\ell \PTm(\cl\clt)\big)
\Big\|_F \leq \sqrt{ \frac{1}{m} \E\Big\|\sum_{\ell=1}^m \varepsilon_\ell \big( \tilde{v}_\ell \PTh(\bl\blt),\tilde{u}_\ell \PTm(\cl\clt)\big)\Big\|_F^2}\\
&\qquad\qquad\qquad = \sqrt{ \frac{1}{m} \sum_{\ell=1}^m \E \Big(\|\tilde{v}_\ell \PTh(\bl\blt)\|_F^2+\|\tilde{u}_\ell \PTm(\cl\clt)\|_F^2\Big)},
\end{align*}
where the last equality follows by going through with the expectation over $\varepsilon_\ell$'s. Recall from the definition of the projection operator that $\PTh(\bl\blt) : = \tfrac{\tilde{\vh}\tilde{\vh}^* }{\|\tilde{\vh}\|_2^2}\bl\blt + \bl\blt \tfrac{\tilde{\vh}\tilde{\vh}^* }{\|\tilde{\vh}\|_2^2}  - \tfrac{\tilde{\vh}\tilde{\vh}^* }{\|\tilde{\vh}\|_2^2}\bl\blt\tfrac{\tilde{\vh}\tilde{\vh}^* }{\|\tilde{\vh}\|_2^2}$, and $\tilde{v}_\ell = |\clt\tilde{\vm}|^2 $. It can be easily verifies that 
$\|\PTh(\bl\blt)\|_F^2 = 2\tfrac{|\blt\tilde{\vh}|^2}{\|\tilde{\vh}\|_2^2} \|\bl\|_2^2 - \tfrac{|\blt\tilde{\vh}|^4}{\|\tilde{\vh}\|_2^4},$
and, therefore, 
\begin{align*}
\E \|\tilde{v}_\ell \PTh(\bl\blt)\|_F^2 &\leq \E |\clt\tilde{\vm}|_2^4 \cdot \E \left(2\tfrac{|\blt\tilde{\vh}|^2}{\|\tilde{\vh}\|_2^2} \|\bl\|_2^2 - \tfrac{|\blt\tilde{\vh}|^4}{\|\tilde{\vh}\|_2^4}\right) \leq 3\|\tilde{\vm}\|_2^4\left(6k-3\right),
\end{align*}
where we used a simple calculation involving fourth moments of Gaussians $\E |\blt\tilde{\vh}|^2 \|\bl\|_2^2 = 3 k \|\tilde{\vh}\|_2^2$. In an exactly similar manner, we can also show that $\|\tilde{u}_\ell \PTm(\cl\clt)\|_F^2 \leq 3\|\tilde{\vh}\|_2^4 (6n-3) $. Putting these together gives us 
\begin{align*}
\E\Big\| \frac{1}{\sqrt{m}}\sum_{\ell=1}^m\varepsilon_\ell \big( \tilde{v}_\ell \PTh(\bl\blt),\tilde{u}_\ell \PTm(\cl\clt)\big)
\Big\|_F \leq 5\max(\|\tilde{\vh}\|_2^2, \|\tilde{\vm}\|_2^2)\sqrt{k+n}. 
\end{align*}
Moreover, 
\begin{align*}
& \E \Big\| \tfrac{1}{\sqrt{m}} \sum_{\ell=1}^m \varepsilon_\ell ( \tilde{v}_\ell\bl\blt,\tilde{u}_\ell \cl\clt)
\Big\| \leq \E\max_{\ell}(\tilde{u}_\ell,\tilde{v}_\ell) \cdot \E\Big\| \tfrac{1}{\sqrt{m}} \sum_{\ell=1}^m \varepsilon_\ell   (\bl\blt,\cl\clt)\Big\|
\end{align*}
Standard net arguments; see, for example, Sec. 5.4.1 of  \cite{eldar2012compressed} show that 
\begin{align*}
\PP \left(\Big\| \tfrac{1}{\sqrt{m}} \sum_{\ell=1}^m \varepsilon_\ell  (\bl\blt,\cl\clt)\Big\| \geq c \sqrt{k+n}\right) \leq \mathrm{e}^{-c m}, ~ \text{provided that} ~ m \geq c (k+n).
\end{align*}
This directly implies that $\E \Big\| \tfrac{1}{\sqrt{m}} \sum_{\ell=1}^m \varepsilon_\ell  (\bl\blt,\cl\clt)\Big\| \leq c\sqrt{k+n}.$ The random variables $u_\ell$ and $v_\ell$ being sub-exponential have Orlicz-1 norms bounded by $c\max(\|\tilde{\vh}\|_2^2,\|\tilde{\vm}\|_2^2)$. Using standard results, such as Lemma 3 in \cite{van2013bernstein}, we then have $\E \max_{\ell} (u_\ell,v_\ell) \leq c\log m.$ Putting these together yields
\begin{align}
\E \Big\| \tfrac{1}{\sqrt{m}} \sum_{\ell=1}^m \varepsilon_\ell ( \tilde{v}_\ell\bl\blt,\tilde{u}_\ell \cl\clt)
\Big\| \leq c\max(\|\tilde{\vh}\|_2^2, \|\tilde{\vm}\|_2^2)\sqrt{(k+n)\log^2 m}.
\end{align}
We have all the ingredients for the final bound on $\mathfrak{C}(\setQ)$ stated below
\begin{align}\label{eq:complexity-estimate}
\mathfrak{C}(\setQ) \leq c\max(\|\tilde{\vh}\|_2^2, \|\tilde{\vm}\|_2^2)\sqrt{(k+n)\log^2 m}. 
\end{align}

\subsection{Probability $p_{\tau}(\setQ)$}
The calculation for the probability $p_{\tau}(\setQ)$, and the positive parameter $\tau$ are given in Appendix due to limitation of space. We find that 
\begin{align}\label{eq:pQ-estimate}
p_{\tau}(\setQ) \geq c > 0, ~ \text{and} ~ \tau = c \max(\|\tilde{\vh}\|_2^2,\|\tilde{\vm}\|_2^2). 
\end{align}
The complexity estimate in \eqref{eq:complexity-estimate}, value of $\tau$ computed above, and $p_{\tau}(\setQ)$ stated in \eqref{eq:pQ-estimate} together with an application of Lemma \ref{lem:Sample-Complexity} prove Theorem \ref{thm:main-theorem}.

\section{Convex Implementation and Phase Transition}
To implement the semi-definite convex program \eqref{eq:convex-optimization-program}, we propose a numerical scheme based on the alternating direction method of multipliers (ADMM). Due to the space limit, the technical details of the algorithm are moved to Section \ref{sec:convopt} of the Appendix.  

\begin{figure}
\centering \begin{overpic}[ height=.33\textwidth,tics=10]{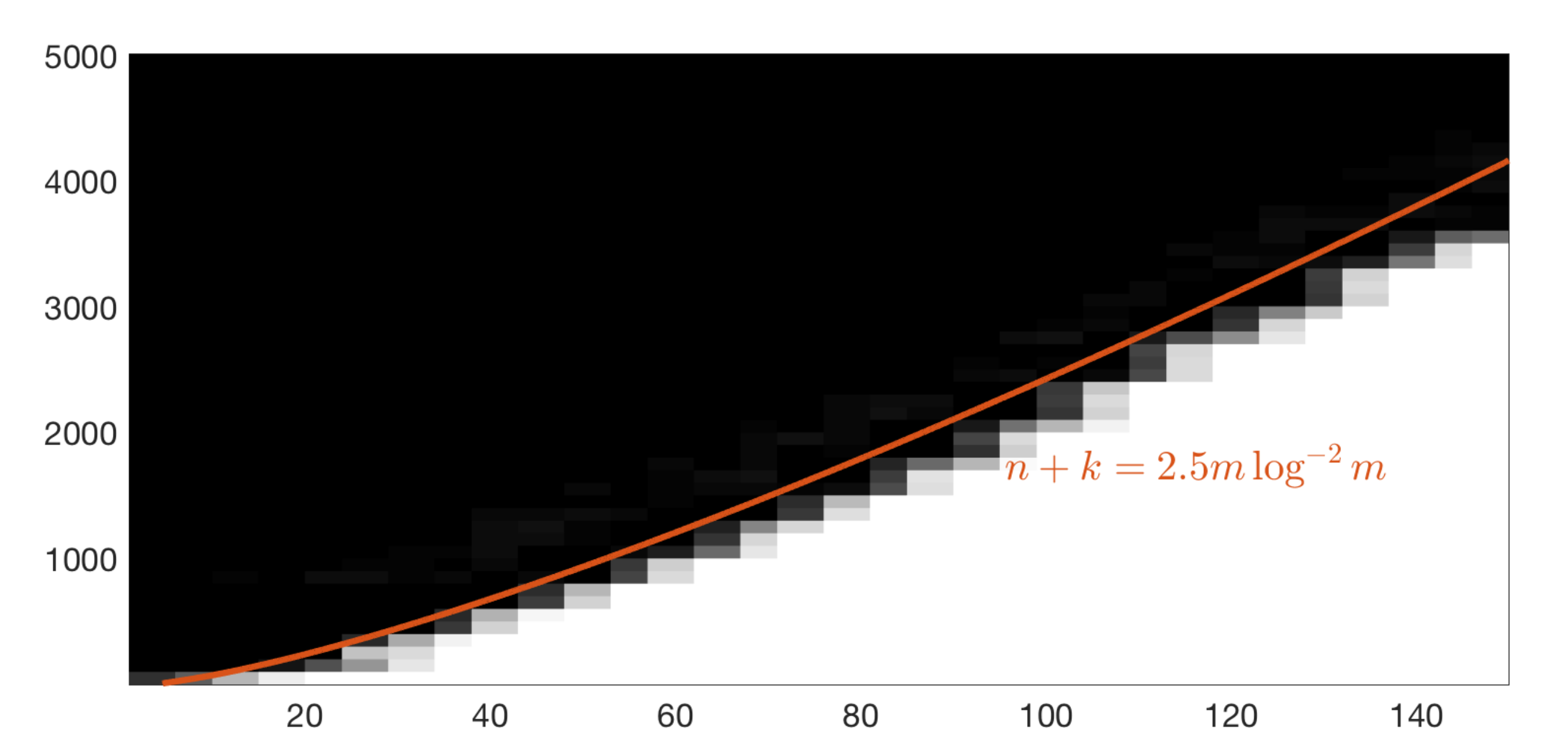}
\put (49,-2.5) {\scalebox{.85}{$n+k$}}
\put (-2,25) {\scalebox{.85}{$m$}}
\end{overpic}\\[.08cm]
 \caption{A phase portrait highlighting the frequency of successful recoveries of the proposed convex program (see the text for the experiment details)}\label{figphase}
\end{figure}

To illustrate the perfect recovery region, in Figure \ref{figphase} we present the phase portrait associated with the proposed convex framework. For each fixed value of $m$, we run the algorithm for 100 different combinations of $n$ and $k$, each time using a different set of Gaussian matrices $\mB$ and $\mC$. If the algorithm converges to a sufficiently close neighborhood of the ground-truth solution (a distance less than 1\% of the solution's $\ell_2$ norm), we label the experiment as successful. Figure \ref{figphase} shows the collected success frequencies, where solid black corresponds to 100\% success and solid white corresponds to 0\% success.  For an empirically selected constant $c$, the success region almost perfectly stands on the left side of the line $n+k = cm\log^{-2}m$.

A similar phase transition diagram can be obtained when $\mB$ is a subset of the columns of identity matrix, and $\mC$ is Gaussian as before. This importantly hints that the convex framework is applicable to more realistic deterministic subspace models. 

\section{Appendix}
The material presented in this section is supplementary to the manuscript above. The dsection contains extended discussions, additional technical proofs and details of the convex program implementation.

\subsection{Visible Light Communication}\label{sec:vis}
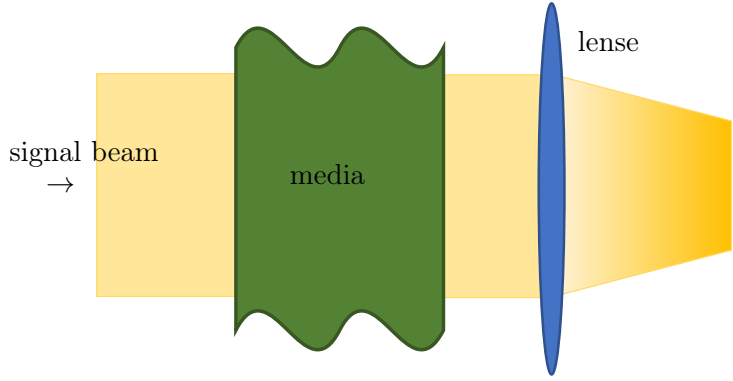
\begin{figure}
\centering
\begin{overpic}[ width=0.57\textwidth,height=0.3\textwidth,tics=10]{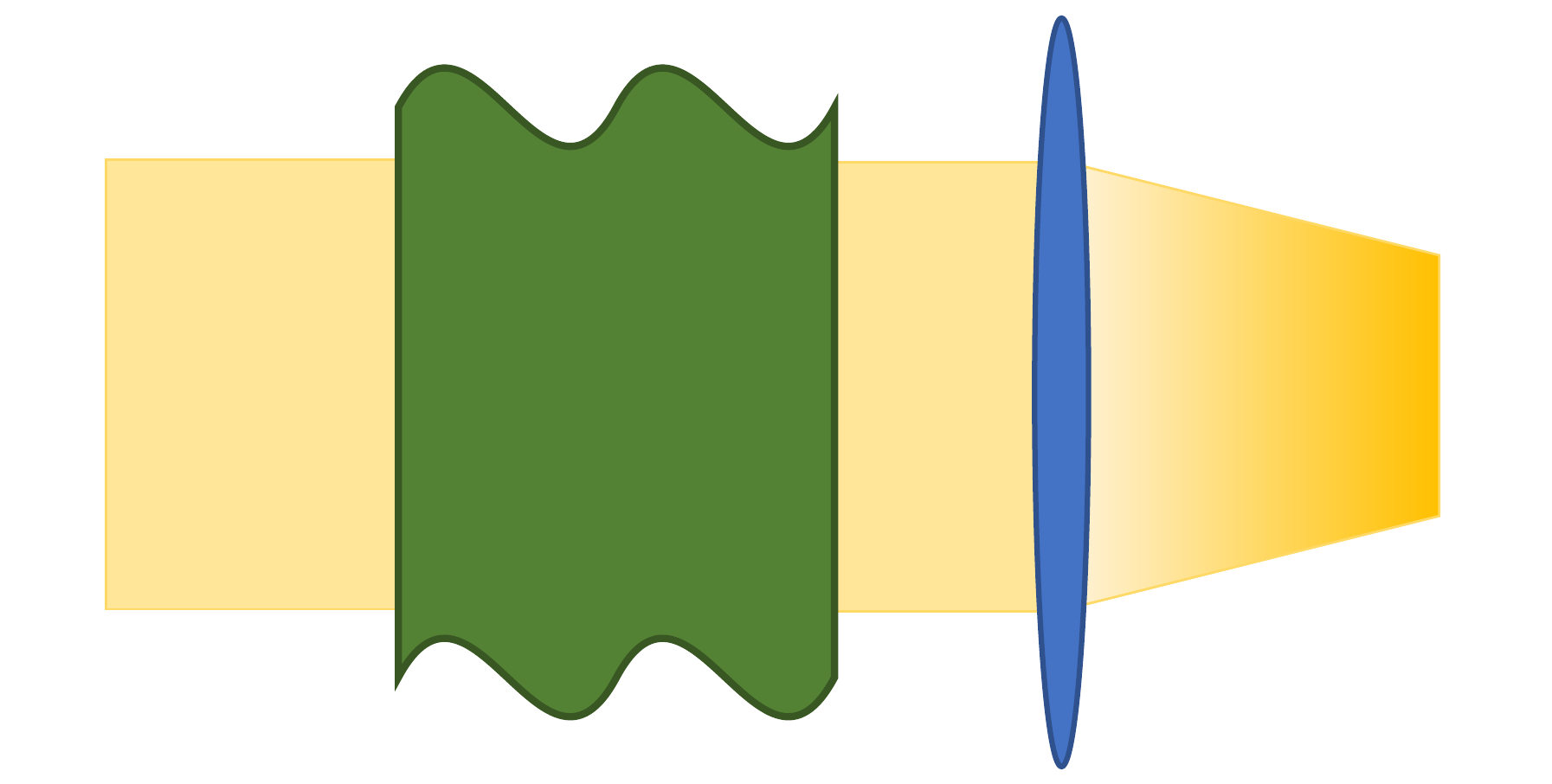}
\linethickness{0.3pt}
\put(0,25.85){$\rightarrow$}
\put(-5,30){signal beam}
\put(72,45){lense}
\put(33,27){media}
 \end{overpic}
 \caption{\small Visible light communication optical setup; the media block normally consists of phosphor, filter and a linear polarizer. The lens takes the Fourier transform of the light and one can only measure the intensity only measurements of this transformed light source signal.}
 \label{fig:VLC}
 \end{figure}

As discussed in the body of the paper, an important application domain where blind deconvolution from phaseless Fourier measurements arises is the visible light communication (VLC). A stylized VLC setup is shown in Figure \ref{fig:VLC}. A message $\vm \in \R^n$ is to be transmitted using visible light. The message is first coded by multiplying it with a tall coding matrix $\mC \in \R^{m \times n}$ and the resultant information $\vx = \mC\vm$ is modulated on a light wave. The light wave propagates through an unknown media. This propagation can be modeled as a convolution $\vx\circledast \vw$ of the information signal $\vx$ with unknown channel $\vw \in \R^m$. The vector $\vw$ contains channel taps, and frequently in realistic applications has only few significant taps. In this case, one can model 
\[
\vw \approx \mB \vh,
\]
where $\vh \in \R^k$ is a short $(k \ll m)$ vector, and $\mB \in \R^{m \times k}$ in this case is a subset of the columns of an identity matrix. Generally, the multipath channels are well modeled with non-zero taps in top locations of $\vw$. In that case, $\mB$ is exactly known to be top few columns of the identity matrix. 

In visible light communication, there is always a difficulty associated with measuring phase information in the received light. Figure \ref{fig:VLC} shows a setup, where we measure the phaseless Fourier transform (light through the lens) of this signal. The measurements are therefore
\[
\vy = |\mF (\mC\vm \circledast \mB\vh)|
\]
and one wants to recover $\vm$, and $\vh$ given the knowledge of $\mB$, and the coding matrix $\mC$. Since we chose $\mC$ to be random Gaussian, and $\mB$ is the columns of identity. As mentioned at the end of the numerics section that with this subspace model, we obtain similar recovery results as one would have for both $\mB$, and $\mC$ being random Gaussians. The proposed convex program solves this difficult inverse problem and recovers the true solution with these subspace models.  

\subsection{Proof of Lemma \ref{lem:Sample-Complexity}}
The proof is based on small ball method developed in \cite{koltchinskii2015bounding,mendelson2014learning} and further studied in  \cite{lecue2018regularization} and \cite{lecue2017regularization}. The proof is mainly repeated using a similar line of argument as in \cite{bahmani2017anchored}, and is provided here for completeness.

Rest of the proof now concerns showing that $(\tilde{\mH},\tilde{\mM})$ is the unique solution to the linearly constrained optimization program \eqref{eq:linearly-constrained-program}. 
Define one sided loss function:
\begin{align}\label{eq:loss}
\setL(\mH,\mM) := \sum_{\ell=1}^m \left(2y^2_\ell - \tfrac{1}{m}\<\bl\blt,\mH\>\<\cl\clt,\tilde{\mM}\> - \< \bl\blt,\tilde{\mH}\>\<\cl\clt,\mM\>  \right)_+,
\end{align}
where $(\cdot)_+$ denotes the positive side. Using this definition, we rewrite \eqref{eq:linearly-constrained-program} compactly as
\begin{align}\label{eq:lin-constr.-loss}
&\minimize~ \|\mH\|_* + \|\mM\|_*\\  
&\text{subject to} \qquad \setL(\mH,\mM) \leq 0\notag.
\end{align}

The goal of the proof is to show that all descent direction $(\dH,\dM) \in \setQ$ that also obey the constraint set have a small $\ell_2$ norm. Since  $(\dH,\dM)$ is a feasible perturbation from the proposed optimal $(\tilde{\mH},\tilde{\mM})$, we have from the constraints above that 
\begin{align}\label{eq:interim}
\setL\big(\tilde{\mH}+\dH,\tilde{\mM}+\dM\big) \leq 0
\end{align}
We begin by expanding the loss function $\text{Loss}(\tilde{\mH}+\dH,\tilde{\mM}+\dM)$ below
\begin{align}\label{eq:interim-eq1}
&\setL(\tilde{\mH}+\dH,\tilde{\mM}+\dM)  \notag\\
&\qquad=  \sum_{\ell=1}^m \Big[(2\yl^2-\big(\<\bl\blt,\tilde{\mH}+\dH\>\<\cl\clt,\tilde{\mM}\> + \<\bl\blt,\tilde{\mH}\>\<\cl\clt,\tilde{\mM}+\dM\>\big)\Big]_+\notag\\
&\qquad= \tfrac{1}{m}\sum_{\ell=1}^m \big[\big(\<\bl\blt,\tilde{\mH}\>\<\cl\clt,\tilde{\mM}\> + \<\bl\blt,\tilde{\mH}\>\<\cl\clt,\tilde{\mM}\>\big) -\notag \\
&\qquad \qquad\qquad \big(\<\bl\blt,\tilde{\mH}+\dH\>\<\cl\clt,\tilde{\mM}\> + \<\bl\blt,\tilde{\mH}\>\<\cl\clt,\tilde{\mM}+\dM\>\big)\Big]_+\notag\\
&\qquad =  \tfrac{1}{m} \sum_{\ell=1}^m \Big[-\<\bl\blt,\dH\> \<\cl\clt,\tilde{\mM}\> -  \<\bl\blt,\tilde{\mH}\>\< \cl\clt,\dM\> \Big]_+\notag\\
&\qquad \qquad \geq\tfrac{1}{m}\sum_{\ell=1}^m\big[(-\big\<\nabla f_\ell,(\dH, \dM)\big\>\big]_+.
\end{align}
where the last equality follows from the using notation $\nabla f_\ell = (\tilde{v}_\ell \bl\blt, \tilde{u}_\ell \cl\clt)$ introduced earlier. Let $\psi_t(s) := (s)_+-(s-t)_+$. Using the fact that $\psi_t(s) \leq (s)_+$, and that for every $\alpha, t \geq 0$, and $s \in \R$, $\psi_{\alpha t}(s) = t\psi_{\alpha}(\frac{s}{t})$, we have
\begin{align}\label{eq:interim-eq2}
&\tfrac{1}{m}\sum_{\ell=1}^m\big[-\big\<\nabla f_\ell,(\dH,\dM)\big\>\big]_+\geq \tfrac{1}{m}\sum_{\ell=1}^m\psi_{\tau \|(\dH,\dM)\|_F}\big\<\nabla f_\ell,(\dH,\dM)\big\>\big]_+\notag\\
& = \|(\dH,\dM)\|_F\cdot \tfrac{1}{m}\sum_{\ell=1}^m\psi_{\tau}\big[-\big\<\nabla f_\ell,\tfrac{(\dH,\dM)}{\|(\dH,\dM)\|_F}\big\>\big]_+\notag\\
&\quad = \|(\dH,\dM)\|_F\Big[\tfrac{1}{m}\sum_{\ell=1}^m\E \psi_{\tau}\big[-\big\<\nabla f_\ell,\tfrac{(\dH,\dM)}{\|(\dH,\dM)\|_F}\big\>\big]_+\notag\\
&- \cdot\tfrac{1}{m}\sum_{\ell=1}^m \big\{ \E \psi_{\tau}\big[-\big\<\nabla f_\ell,\tfrac{(\dH,\dM)}{\|(\dH,\dM)\|_F}\big\>\>\big]_+- \psi_{\tau}\big[-\big\<\nabla f_\ell,\tfrac{(\dH,\dM)}{\|(\dH,\dM)\|_F}\big\>\big]_+ \big\}\Big].
\end{align}
Define a centered random process $\mathcal{R}(\mB,\mC)$ as follows
\begin{align*}
&\mathcal{R}(\mB,\mC):=\\
&\sup_{(\dH,\dM)\in \setQ}\tfrac{1}{m}\sum_{\ell=1}^m\Big(\E \psi_{\tau}\big[-\big\<\nabla f_\ell,\tfrac{(\dH,\dM)}{\|(\dH,\dM)\|_F}\big\>\>\big]_+- \psi_{\tau}\big[-\big\<\nabla f_\ell(\tilde{\mH},\tilde{\mM}),\tfrac{(\dH,\dM)}{\|(\dH,\dM)\|_F}\big\>\big]_+ \Big)
\end{align*}
and an application of bounded difference inequality \cite{mcdiarmid1989method} yields that $\mathcal{R}(\mB,\mC) \leq \E \mathcal{R}(\mB,\mC) + t\tau/\sqrt{m}$  with probability at least $1-\mathrm{e}^{-2mt^2}$. It remains to evaluate $\E  \mathcal{R}(\mB,\mC)$, which after using a simple symmetrization inequality \cite{van1997weak} yields 
\begin{align}
&\E \setR(\mB,\mC) \leq 2\E \sup_{(\dH,\dM)\in \setQ}\tfrac{1}{m}\sum_{\ell=1}^m\varepsilon_\ell \psi_{\tau}\big[-\big\<\nabla f_\ell,\tfrac{(\dH,\dM)}{\|(\dH,\dM)\|_F}\big\>\big]_+,
\end{align}
where $\varepsilon_1, \varepsilon_2, \ldots, \varepsilon_m$ are independent Rademacher random variables. Using the fact that $\psi_t(s)$ is a contraction: $|\psi_t(\alpha_1)-\psi_t(\alpha_2)| \leq |\alpha_1-\alpha_2|$ for all $\alpha_1, \alpha_2 \in \R$, we have from the Rademacher contraction inequality \cite{ledoux2013probability} that 
\begin{align}\label{eq:random-process}
&\E \sup_{(\dH,\dM)\in \setQ}\tfrac{1}{m}\sum_{\ell=1}^m\varepsilon_\ell \psi_{\tau}\big[-\big\<\nabla f_\ell,\tfrac{(\dH,\dM)}{\|(\dH,\dM)\|_F}\big\>\big]_+\leq \E \sup_{(\dH,\dM)\in \setQ}\tfrac{1}{m}\sum_{\ell=1}^m-\varepsilon_\ell\big\<\nabla f_\ell,\tfrac{(\dH,\dM)}{\|(\dH,\dM)\|_F}\big\>\notag\\
&\qquad\qquad = \E \sup_{(\dH,\dM)\in \setQ}\tfrac{1}{m}\sum_{\ell=1}^m\varepsilon_\ell\big\<\nabla f_\ell,\tfrac{(\dH,\dM)}{\|(\dH,\dM)\|_F}\big\>, 
\end{align}
where the last equality is the result of the fact that a global sign change of a sequence of  Rademacher random variables does not change their distribution. In addition, using the facts that $t\mathbf{1}(s\geq t) \leq \psi_t(s)$, and that random vectors $\nabla f_1, \nabla f_2, \ldots, \nabla f_m$ are identically distributed and the distribution is symmetric, it follows 
\begin{align}\label{eq:tail-prob}
\tau\PP\big(\big\<\nabla f_\ell,\tfrac{(\dH,\dM)}{\|(\dH,\dM)\|_F}\big\>\geq \tau\big) &= \tau\E \big(\mathbf{1}\big[{ \big\<\nabla f_\ell,\tfrac{(\dH,\dM)}{\|(\dH,\dM)\|_F}\big\>\geq \tau}\big]\big)\notag\\
&\leq \E \psi_{\tau}\left[ \big\<\nabla f_\ell,\tfrac{(\dH,\dM)}{\|(\dH,\dM)\|_F}\big\>\right]. 
\end{align}
Plugging \eqref{eq:tail-prob}, and \eqref{eq:random-process} in \eqref{eq:interim-eq2}, we have 
\begin{align*}
&\tfrac{1}{m}\sum_{\ell=1}^m\big[-\big\<\nabla f_\ell,\tfrac{(\dH,\dM)}{\|(\dH,\dM)\|_F}\big\>\big]_+\geq \tau\|(\dH,\dM)\|_F\cdot\PP\big(\big\<\nabla f_\ell,\tfrac{(\dH,\dM)}{\|(\dH,\dM)\|_F}\big\>\geq \tau\big)  \\
&-2\|(\dH,\dM)\|_F \E \sup_{(\dH,\dM)\in \setQ} \tfrac{1}{m}\sum_{\ell=1}^m\varepsilon_\ell\big\<\nabla f_\ell,\tfrac{(\dH,\dM)}{\|(\dH,\dM)\|_F}\big\>-2\|(\dH,\dM)\|_F\tfrac{t\tau}{\sqrt{m}}.\\
\end{align*}
Combining this with \eqref{eq:interim} and \eqref{eq:interim-eq1}, we obtain the final result 
\begin{align*}
&\tau\|(\dH,\dM)\|_F\Big[\PP\big(\big\<\nabla f_\ell,\tfrac{(\dH,\dM)}{\|(\dH,\dM)\|_F}\big\>\geq \tau\big)-2\E \sup_{(\dH,\dM)\in \setQ} \tfrac{1}{m}\sum_{\ell=1}^m\varepsilon_\ell\big\<\nabla f_\ell,\tfrac{(\dH,\dM)}{\|(\dH,\dM)\|_F}\big\>\Big]\\
&\qquad\qquad -2\|(\dH,\dM)\|_F\tfrac{t\tau}{\sqrt{m}} \leq 0.
\end{align*}
Using the definitions in \eqref{eq:Rademacher-Complexity}, and \eqref{eq:p_tau}, we can write 
\begin{align*}
\|(\dH,\dM)\|_F \left(\tau p_{\tau}(\setQ)- \frac{(2\mathfrak{C}(\setQ) + t\tau)}{\sqrt{m}}\right) \leq 0.
\end{align*}
It is clear that  choosing $m \geq \left( \frac{2\mathfrak{C}(\setQ)+t\tau}{\tau p_\tau(\setQ)}\right)^2$ implies 
\begin{align*}
(\dH,\dM) = (\mathbf{0},\mathbf{0}).
\end{align*}
The proof is complete.

\subsection{Probability $p_{\tau}(\setQ)$}
In this section, we determine the probability $p_{\tau}(\setQ)$, and the positive parameter $\tau$ in \eqref{eq:p_tau} for the set $\setQ$ in \eqref{eq:setQ}. For a point $(\dH,\dM) \in \setQ$, and randomly chosen $\nabla f_\ell$, we have via Paley Zygmund inequality that 
\begin{align*}
&\mathbb{P}\Big( \left|\left\<\nabla f_\ell ,(\dH,\dM)\right\> \right|^2 \geq \frac{1}{2} \E\left|\left\<\nabla f_\ell,(\dH,\dM)\right\>  \right|^2 \Big)\geq \frac{1}{4} \frac{\big(\E\left| \left\<\nabla f_\ell ,(\dH,\dM)\right\>  \right|^2\big)^2}{\E\left| \left\<\nabla f_\ell,(\dH,\dM)\right\>  \right|^4}.
\end{align*}
The particular choice of random gradient vectors we are using is $\nabla f_\ell = (\tilde{v}_\ell \bl\blt, \tilde{u}_\ell \cl\clt)$ giving us $\left\<\nabla f_\ell ,(\dH,\dM)\right\> = \tilde{v}_\ell\<\bl\blt, \dH\> + \tilde{u}_\ell \<\cl\clt,\dM\>$. Since $\bl$, and $\cl$ are standard Gaussian vectors, using the equivalence of $L_p$-norms for Gaussians, we deduce that
\[
\left(\E\left| \tilde{v}_\ell\<\bl\blt, \dH\> + \tilde{u}_\ell \<\cl\clt,\dM\> \right|^4\right)^{1/4}  \leq c \left(\E\left| \tilde{v}_\ell\<\bl\blt, \dH\> + \tilde{u}_\ell \<\cl\clt,\dM\> \right|^2\right)^{1/2}.
\]
Plugging last two inequalities in \eqref{eq:p_tau} reveals that
\begin{align}\label{eq:supplementary-pQ-estimate}
p_{\tau}(\setQ) \geq c > 0
\end{align}
for an absolute constant $c$. To compute $\tau$, we expand $\E\left|\left\<\nabla f_\ell,(\dH,\dM)\right\>  \right|^2$ giving us
\begin{align}\label{eq:tau-estimate-1}
&\E\left| \tilde{v}_\ell\<\bl\blt, \dH\> + \tilde{u}_\ell \<\cl\clt,\dM\> \right|^2 
= 3\|\tilde{\vm}\|_2^4 (\<\text{diag}(\dH),\dH\> + 2\|\dH\|_F^2) \notag\\
&\qquad\qquad +3 \|\tilde{\vh}\|_2^4(\<\text{diag}(\dM),\dM\> + 2\|\dM\|_F^2) + 2|\tilde{\vh}^*\text{diag}(\dH)\tilde{\vh} + 2\tilde{\vh}^*\dH\tilde{\vh}|^2,
\end{align}
where we have made use of multiple simple facts including that $\E |\tilde{u}_\ell|^2 = 3 \|\tilde{\vh}\|_2^4$, and similarly for $\tilde{v}_\ell$, and two identities: $
\E|\blt\tilde{\vh}|^2 \blt\dH\bl = \tilde{\vh}^* \text{diag}(\dH)\tilde{\vh} + 2\tilde{\vh}^*\dH\tilde{\vh},$
and $\E(\blt\dH\bl) \bl\blt = \text{diag}(\dH) + 2(\dH) \implies \E|\blt\dH\bl|^2 = \<\text{diag}(\dH),\dH\> + 2\|\dH\|_F^2.$
We also made use of the fact that $\setQ \perp \setN$ and therefore $\<\tilde{\mH},\dH\>-\<\tilde{\mM},\dM\> = 0 $, or equivalently, $\tilde{\vh}^*\dH\tilde{\vh} = \tilde{\vm}^*\dM\tilde{\vm}$. 

It is easy to conclude from \eqref{eq:tau-estimate-1} now that 
\begin{align*}
&\E\left| \tilde{v}_\ell\<\bl\blt, \dH\> + \tilde{u}_\ell \<\cl\clt,\dM\> \right|^2  
\geq 6(\|\tilde{\vh}\|_2^4 \|\dH\|_F^2 + \|\tilde{\vm}\|_2^4 \|\dM\|_F^2).
\end{align*}
This directly means, we can take $\tau = c \max(\|\tilde{\vh}\|_2^2,\|\tilde{\vm}\|_2^2)$, where $c$ is an absolute constant. 


\subsection{Implementing the Convex Program}\label{sec:convopt}
In this section we take an alternating direction method of multipliers (ADMM) scheme to address \eqref{eq:convex-optimization-program}, which takes the form
\begin{align}\label{eq:orig}
&\underset{\mX_1,\mX_2}{\text{minimize}}~ \trace(\mX_1) + \trace(\mX_2)\\
&\text{subject to} ~ \left\< \ba{1}\ba{1}^*,\mX_1\right\>\left\< \ba{2}\ba{2}^*,\mX_2\right\> \geq  \delta_\ell\geq 0, \notag \\
& \qquad\qquad \qquad ~ \ell = 1,2, \ldots, L,\notag \\
& \qquad\qquad ~~~ \mX_1\succcurlyeq \boldsymbol{0},  ~\mX_2 \succcurlyeq \boldsymbol{0}. \notag 
\end{align}
Note that for a complex matrix $\mX$ being Hermitian is a requirement for being positive semidefinite. For a simpler notation we define the convex set
\begin{equation}\label{eq:C}
\mathcal{C} = \left\{ \left(\vu,\vv \right)\in\mathbb{R}^L\times\mathbb{R}^L: u_\ell v_\ell\geq \delta_\ell>0,u_\ell\geq 0 \right\}.
\end{equation}
In order to derive the ADMM scheme, after introducing new variables, program \eqref{eq:orig} can be written as 
\begin{align}\label{eq:cast}
&\underset{\{\mX_i,\mZ_i,\vu_i\}_{i=1,2}}{\text{minimize}}~ \ind_{\mathcal{C}}(\vu_1,\vu_2) + \sum_{j=1}^2\trace(\mX_j) + \ind_+(\mZ_j)\\
&\text{subject to} ~~~~ u_{j,\ell} = \left\< \ba{j}\ba{j}^*,\mX_j\right\>, \notag  ~ \ell = 1,2, \ldots, L\notag, ~ j=1,2, \\
& \qquad\qquad ~~~~~ \mX_j = \mZ_j, ~ j=1,2, \notag 
\end{align}
where the constraints are reflected in the indicator functions
\[\ind_{\mathcal{C}}(\vu,\vv) = \left\{\begin{array}{lc}0 & (\vu,\vv)\in\mathcal{C}\\ +\infty& (\vu,\vv)\notin\mathcal{C}\end{array}  \right., ~~~ \ind_+(\mZ) = \left\{\begin{array}{lc}0 & \mZ\succeq \boldsymbol{0}\\ +\infty& \mZ\nsucceq \boldsymbol{0}\end{array}  \right..
\] 
Defining the dual matrices $\mP_1, \mP_2$ and the dual vectors $\balpha_1, \balpha_2\in \mathbb{R}^L$, the augmented Lagrangian for \eqref{eq:cast} takes the form 
\begin{align}L\left( \{\mX_i,\mZ_i,\mP_i,\vu_i,\balpha_i\}_{i=1,2} \right) &=  \ind_{\mathcal{C}}(\vu_1,\vu_2) + \sum_{j=1}^2\trace(\mX_j) + \ind_+(\mZ_j) \notag \\ & ~~~~+ \frac{\rho_1}{2} \sum_{j=1}^2\sum_{\ell=1}^L \left( u_{j,\ell} - \left\< \ba{j}\ba{j}^*,\mX_j \right\>+ \alpha_{j,\ell}\right)^2 \notag\\ & ~~~~+ \frac{\rho_2}{2}\sum_{j=1}^2 \left\|\mX_j -\mZ_j + \mP_j \right\|_F^2.
\end{align}
In an ADMM scheme the update for each variable at the $k$-th iteration is performed by minimizing $L$ with respect to that variable, while fixing the other ones. More specifically, using the superscript $(k)$ to denote the iteration, for $j=1,2$ we have the primal updates 
\begin{align*}
&\mX_j^{(k+1)} = \argmin_{\mX_j}  \trace(\mX_j) + \frac{\rho_1}{2} \!\sum_{\ell=1}^L\! \left( \left\< \ba{j}\ba{j}^*,\mX_j \right\> - u_{j,\ell}^{(k)} -  \alpha_{j,\ell}^{(k)}\right)^2 \!+ \!\frac{\rho_2}{2} \left\|\mX_j -\mZ_j^{(k)} + \mP_j^{(k)} \right\|_F^2,\\ &\mZ_j^{(k+1)} = \argmin_{\mZ_j}~ \frac{1}{2}\left\|\mZ_j - \mX_j^{(k+1)} -  \mP_j^{(k)} \right\|_F^2 + \ind_+(\mZ_j),\\ &\left(\vu_1^{(k+1)},\vu_2^{(k+1)}\right) = \argmin_{\vu_1,\vu_2} ~ \frac{1}{2}\sum_{j=1}^2\sum_{\ell=1}^L \left( u_{j,\ell} - \left\< \ba{j}\ba{j}^*,\mX_j^{(k+1)} \right\>+ \alpha_{j,\ell}^{(k)}\right)^2 + \ind_{\mathcal{C}}(\vu_1,\vu_2),
\end{align*}
and the dual updates 
\begin{align*}
 \alpha_{j,\ell}^{(k+1)} &= \alpha_{j,\ell}^{(k)} + u_{j,\ell}^{(k+1)} - \left\< \ba{j}\ba{j}^*,\mX_j^{(k+1)} \right\>\\ \mP_j^{(k+1)} & = \mP_j^{(k)} + \mX_j^{(k+1)} -\mZ_j^{(k+1)}.  
\end{align*}
In the sequel we derive closed-form expressions for all the primal updates. To formulate the $\mX$-update, taking the derivative of the objective with respect to $\mX_j$ and setting it to zero yields
\[\mI + \rho_1 \sum_{\ell=1}^L\left( \left\< \ba{j}\ba{j}^*,\mX_j^{(k+1)} \right\> - u_{j,\ell}^{(k)} -  \alpha_{j,\ell}^{(k)}\right)\ba{j}\ba{j}^* + \rho_2\left( \mX_j^{(k+1)} -\mZ_j^{(k)} + \mP_j^{(k)} \right) = \boldsymbol{0},
\] 
which after vectorizing $\mX_j^{(k+1)}$  yields 
\[\mbox{vec}\!\left( \mX_j^{(k+1)} \right)  = \mA_j^{-1}~\mbox{vec}\left(\rho_1\sum_{\ell=1}^L\left( u_{j,\ell}^{(k)} +   \alpha_{j,\ell}^{(k)} \right) \ba{j}\ba{j}^*  + \rho_2\left( \mZ_j^{(k)} - \mP_j^{(k)} \right)-\mI \right),
\]
where
\[\mA_j = \rho_1\sum_{\ell=1}^L \mbox{vec}\left( \ba{j}\ba{j}^*\right)\mbox{vec}\left( \ba{j}\ba{j}^*\right)^* + \rho_2 \mI.
\]
Note that $\mA_j^{-1}$ only needs to be calculated once throughout the entire process.  

The $\mZ$ update is basically the projection of a Hermitian matrix onto the PSD cone. Considering the eigen-decomposition of the Hermitian matrix $\tilde \mZ\in \mathbb{C}^{n\times n}$:
\[\tilde\mZ = \mU\mbox{diag}\left(\lambda_1,\cdots,\lambda_n\right)\mU^*,
\]
all eigenvalues are real, and the solution to 
\[\underset{\mZ}{\text{minimize}}~~\frac{1}{2}\left\| \mZ - \tilde\mZ\right\|_F^2+ \ind_+(\mZ)
\]
is simply $\mU\mbox{diag}\left(\max(\lambda_1,0),\cdots,\max(\lambda_n,0)\right)\mU^*$. 

Finally, the $\vu$-update step in the proposed ADMM scheme requires a fast formulation of the projection onto the set $\mathcal{C}$. It is straightforward to see that program
\begin{equation}\underset{\vu_1,\vu_2}{\text{minimize}} ~ \frac{1}{2}\sum_{j=1}^2\sum_{\ell=1}^L \left( u_{j,\ell} - \xi_{j,\ell}\right)^2 + \ind_{\mathcal{C}}(\vu_1,\vu_2) \label{eq:projhyp}
\end{equation}
decouples into $L$ distinct programs of the form 
\begin{equation}\underset{u_1,u_2}{\text{minimize}}  ~ \frac{1}{2}\sum_{j=1}^2\left( u_{j} - \xi_{j}\right)^2 ~~ \mbox{subject to:}~~ u_1u_2\geq \delta> 0, ~ u_1\geq 0. \label{eq:projhypdecoup}
\end{equation}
Note that since the case $u_1u_2=0$ leads to a trivial argument, we consider the strict inequality $\delta>0$. In the sequel we focus on addressing \eqref{eq:projhypdecoup}, as solving \eqref{eq:projhypdecoup} for each component $\ell$ would deliver the solution to \eqref{eq:projhyp}. We proceed by forming the Lagrangian for the constrained problem \eqref{eq:projhypdecoup}
\[l(u_1,u_2,\mu_1,\mu_2) = \frac{1}{2}\left\|\begin{pmatrix}u_1\\ u_2 \end{pmatrix} -  \begin{pmatrix} \xi_1\\ \xi_2  \end{pmatrix}\right\|^2 + \mu_1\left( \delta - u_1 u_2 \right ) - \mu_2u_1.
\]
Along with the primal constraints, the Karush-Kuhn-Tucker optimality conditions are 
\begin{align}
\label{e5}
\frac{\partial l}{\partial u_1} = u_1 - \xi_1 - \mu_1 u_2  - \mu_2 &=0,\\ 
\label{e6}
\frac{\partial l}{\partial u_2} = u_2 - \xi_2 - \mu_1 u_1   &=0,\\ 
\notag 
\mu_1\geq 0, \quad \mu_1\left( \delta - u_1 u_2 \right ) &=0,\\
\notag \mu_2 \geq 0, \quad \mu_2u_1&=0.
\end{align}
We now proceed with the possible cases.

\textbf{Case 1.} $\mu_1=\mu_2=0$:\\
In this case we have $(u_1,u_2)=(\xi_1,\xi_2)$ and this result would only be acceptable when $u_1u_2 \geq \delta$ and $u_1\geq 0$.

\textbf{Case 2.} $\mu_1=0$, $u_1 =0$:\\ 
In this case the first feasibility constraint of \eqref{eq:projhypdecoup} requires that $\delta\leq 0$, which is not a possiblity.

\textbf{Case 3.} $\delta - u_1 u_2 = 0$, $u_1 =0$:\\ 
Similar to the previous case, this cannot happen when $\delta>0$.

\textbf{Case 4.} $\mu_2=0$, $\delta - u_1 u_2  =0$:\\ 
In this case we have $\delta = u_1 u_2$, combining which with \eqref{e6} yields $\delta = (\xi_2 + \mu_1 u_1)u_1$, or
\begin{align}\label{e17}
\mu_1 = \frac{\delta - \xi_2u_1}{u_1^2}.
\end{align}
Similarly, \eqref{e5} yields 
\begin{equation}\label{e19}
u_1 = \xi_1 + \mu_1(\xi_2 + \mu_1 u_1).
\end{equation}
Since the condition $\delta=u_1u_2$ requires that $u_1>0$, $\mu_1$ can be eliminated between \eqref{e17} and \eqref{e19} to generate the following fourth order polynomial equation in terms of $u_1$:
\begin{align*}
u_1^4-\xi_1 u_1^3 +\delta\xi_2 u_1- \delta^2=0.
\end{align*}
After solving this 4-th order polynomial equation, we pick the real root $u_1$ which obeys
\begin{align}\label{eqconsts}
u_1\geq 0, \qquad \delta - \xi_2u_1\geq 0.
\end{align}
Note that the second inequality in \eqref{eqconsts} warrants nonnegative values for $\mu_1$ thanks to \eqref{e17}. After picking the right root, we can explicitly obtain $\mu_1$ using \eqref{e19} and calculate the $u_2$ using \eqref{e6}. The resulting $(u_1,u_2)$ pair presents the solution to \eqref{eq:projhypdecoup}, and finding such pair for every $\ell$ provides the solution to \eqref{eq:projhyp}.


\newcommand{\etalchar}[1]{$^{#1}$}

\end{document}